\pdfoutput=1

\documentclass[twocolumn,showpacs]{revtex4}
\usepackage{amssymb,amsmath,graphicx}
\usepackage{color}
\usepackage[T1]{fontenc}
\begin{document}

\title{
%  Pairwise entanglement, charge-spin separation, and the Mott transition
%  for correlated electrons in nanochains
%
  Pairwise entanglement and the Mott transition
  for correlated electrons in nanochains
}

\author{Adam Rycerz}
\affiliation{Marian Smoluchowski Institute of Physics, 
Jagiellonian University, {\L}ojasiewicza 11, PL--30348 Krak\'{o}w, Poland}

\begin{abstract}
Pairwise entanglement, calculated separately for charge and spin degrees
of freedom, is proposed as a ground-state signature of the Mott transition
in correlated nanoscopic systems. Utilizing the exact diagonalization -- 
ab initio method (EDABI), for chains containing $N\leqslant{}16$ 
hydrogenic-like atoms (at the half filling), we find that the vanishing of the
nearest-neighbor charge concurrence indicates the crossover from 
a~partly-localized quantum liquid to the Mott insulator. Spin concurrence 
remains nonzero at the insulating phase, showing that the decopling of spin 
and charge degrees of freedom may manifest itself by wavefunctions entangled 
in spin, but separable in charge coordinates. 
At the quarter filling, the analysis for $N\leqslant{}20$ 
shows that spin concurrence vanishes immediately when the 
charge-energy gap obtained from the scaling with $1/N\rightarrow{}0$ vanishes, 
constituting a~finite-system version of the Mott transition.
Analytic derivations of the formulas expressing either charge or spin 
concurrence in terms of ground-state correlation functions are also provided. 
\end{abstract}

%\date{\today}
\date{March 25, 2017}
\maketitle

%%%%%%%%%%%%%%%%%%%%%%%%%%%%%%%% INTRODUCTION %%%%%%%%%%%%%%%%%%%%%%%%%%%%%%%%%%

\section{Introduction}
In 90 years after the seminal work by Schr\"{o}dinger \cite{Sch26} analytical properties of quantum-mechanical wavefunctions describing hydrogenic-like atoms (or ions) continue to surprise. For instance, recent derivation of the Wallis formula for $\pi$ by Friedmann and Hagen \cite{Fri15} refers to the variational principle and the correspondence principle, but also employs a~counterintuitive fact that some variational approaches may truncate particular excited states of atoms with the same or better accuracy than the ground state (GS). This feature has direct analogs in many-electron hydrogenic systems and motivated the proposal of the Exact Diagonalization  -- Ab Initio method (EDABI) \cite{Ryc04}, which were recently used to discuss the influence of electron correlations on the metallization of solid hydrogen in a~rigorous manner \cite{Kad14,Kad15}. Also, as a~method putting equal footing on single- and multiparticle aspects of the correlated quantum states, EDABI seems to be a~promising candidate for the theoretical tool capable of giving a~better insight into the superconductivity mechanism in sulfur hydrides \cite{Dro15,Err15} or into the magnetic moment formation in functionalized graphenes \cite{Yaz10}. 

When considering a~generic second-quantized Hamiltonian with both spin and charge degrees of freedom, a~few numerical techniques can be regarded as exact ones; i.e., giving the desired correlation functions within the accuracy limited (in principle) by the machine precision only. These includes: Exact Diagonalization (ED) for relatively small systems \cite{Pre13}, Density Matrix Renormalization Group (DMRG) for low-dimensional systems \cite{DeCh08}, and Quantum Monte Carlo (QMC) for non-frustrated systems \cite{Sug86}. A~separate class is outlined by variational approaches designed to treat one-dimensional (1D) atomic chains in the insulating phase \cite{Ste11}. Even though each of these methods provides us with detailed information about the closed-system GS, it is usually a~challenging task to determine whether GS is insulating, metallic, or of a~more complex nature \cite{spnlqd}. This is primarily because standard GS signatures of the metal-insulator transition, such as a~vanishing charge gap \cite{Hir85}, are absent at any finite system size (quantified by the number of lattice sites $N$), but may appear only after the so-called finite-size scaling (with $1/N\rightarrow{}0$), a~procedure introducing systematic errors difficult to estimate in some cases. 

For this reason, specially-designed GS correlation functions, not only signalling the Mott transition in the thermodynamic limit, but also showing fast convergence (with $1/N\rightarrow{}0$) for finite systems close to the metal-insulator boundary, may constitute a~valuable complement of the existing numerical techniques. 

In attempt to propose such correlation functions, one need to point out the relevance of quantum fluctuations between electronic doubly-occupied sites and unoccupied sites and between singly occupied sites with spin up and spin down in lattice models, as suggested before by numerous authors \cite{Kap82,Yok90,Str99,Yok11,Zho14}. The concept quantum entanglement \cite{Ein35,Ben00}, together with entanglement measures such as the concurrence \cite{Woo98}, was adopted to quantify the above-mentioned fluctuations in various systems \cite{Ost02,Ver04,Wez05,Sch01,Zan02,SGu04,Wie08,Bog12,Gru15,Ole06,Ole12,Che15,You15,Ryc06d,Ram06,Ryc08a,Too14,Mot14,Sch13}, including basic spin models \cite{Ost02,Ver04,Wez05,Hua14}, fermionic systems \cite{Sch01,Zan02,SGu04,Wie08,Bog12,Gru15}, or systems with spin and orbital degrees of freedom \cite{Ole06,Ole12,Che15,You15}. 
For an open fermionic system (namely: a~correlated quantum dot attached to the leads) we found that the concurrence, defined separately for charge and spin degrees of freedom, allowed one to distinguish between different quantum transport regimes of the system \cite{Ryc06d}. Analogs of this observation were also reported for double \cite{Ram06,Ryc08a} and triple quantum dots \cite{Too14}. For closed systems, in particular for small molecules, Mottet {\em et al.}\ \cite{Mot14} showed that the entanglement analysis provided a valuable insight into the chemical bond formation. Even noninteracting systems with complex Fermi surface topology were characterized via their entanglement spectra \cite{Sch13}. Also very recently, a~generic variational approach to correlated quantum systems, in which the output numerical precision is steered by setting maximal allowed entanglement between a~selected subsystem and the environment, was proposed \cite{Cza15}. 

Sacramento {\em et al.}\ \cite{Sac13} used entanglement measures to complement the long-lasting discussion of decoupling of charge and spin degrees of freedom in 1D Hubbard model \cite{Lie68} and its relation to the charge-spin separation phenomenon in Luttinger liquids \cite{Voi93,Ber00}. In particular, it is pointed out in Ref.\ \cite{Sac13} that charge-charge fluctuations, when quantified by properly defined correlation functions, show substantially different asymptotic behavior than spin-spin fluctuations in both the metallic and the insulating phases. 

\begin{figure}
\centerline{\includegraphics[width=0.7\linewidth]{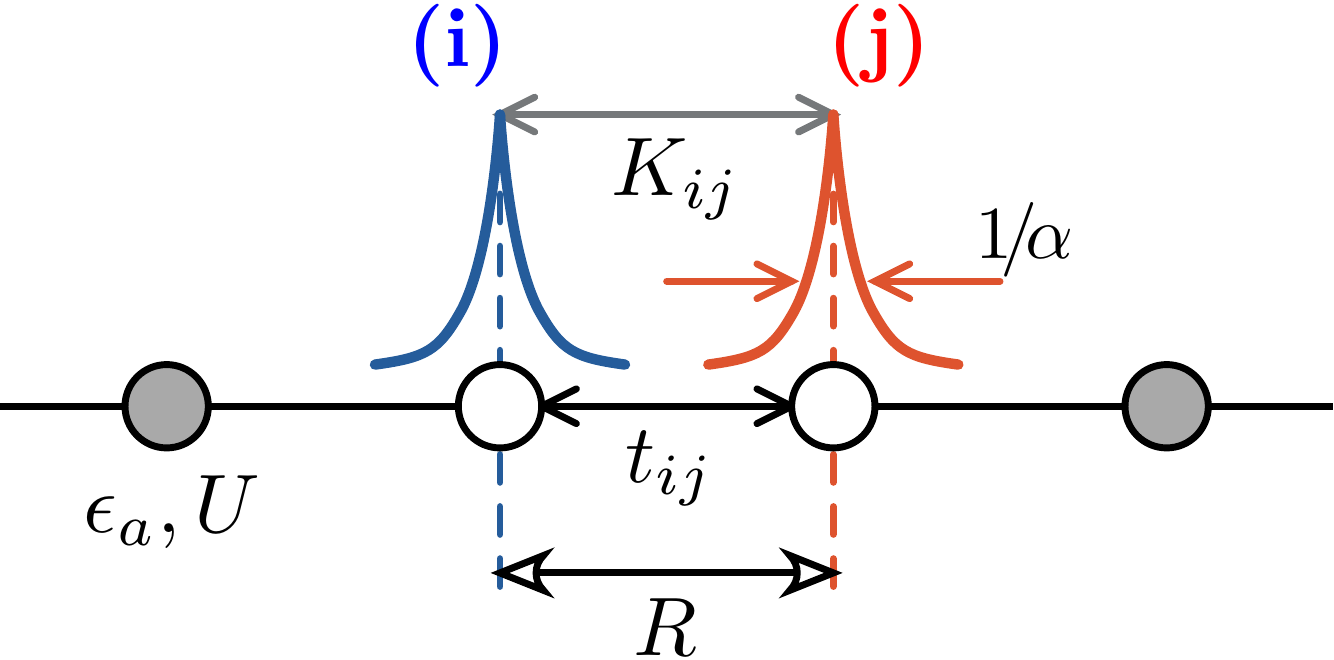}}
\caption{ \label{fig:chain}
  A system studied numerically (schematic). Hydrogenic-like atoms, containing 1$s$-type orbitals with their radii $\alpha^{-1}$, are arranged linearly with the interatomic distance $R$. The Hamiltonian parameters, including the atomic energy $\epsilon_a$, the intraatomic Coulomb integral $U$, the hopping integrals $t_{ij}$, and interatomic Coulomb integrals $K_{ij}$, are determined for each value of  $\alpha$ and $R$ (see Sec.\ II for the details). 
}
\end{figure}

Here we focus on linear chains containing up to $N=20$ equally-spaced hydrogenic-like atoms, each one containing a~single valence orbital (see Fig.\ \ref{fig:chain}). Although such chains are not directly observable due to the well-known Peierls instability, they have been intensively studied to benchmark different {\em ab initio} approaches to strongly-correlated systems \cite{Hach06}. 
In this paper, two distinct physical regimes are considered: At the half filling the system shows a crossover behavior, with the increasing interatomic distance $R$, from a~{\em partly localized} quantum liquid to the Mott insulating phase \cite{Ryc04,Spa01}. At the quarter filling, the Mott transition is reconstructed with $1/N\rightarrow{}0$ \cite{Ryc04}. (The above-mentioned findings are consistent with more recent results of QMC simulations for Hubbard chains with long-range interactions \cite{Hoh12}.) 
It is worth to mention that the two regimes are also significantly different from a~more fundamental point of view: in the strong-coupling (i.e., large $R$) limit the charge fluctuations are suppressed and the half-filled chain can be effectively described as 1D Heisenberg antifferomagnet, for which long-range order is absent due to the Mermin-Wagner theorem \cite{Mer66}. For the quarter-filling similar reasoning cannot be applied in the presence of long-range Coulomb interactions, and the charge-density wave phase is predicted to appear \cite{Ryc04,Eji05,Sch90,Mey09}.
In other words, we consider the two complementary model cases allowing one to test entanglement-based phase-transition indicators. We further discuss the concurrence \cite{Woo98}, defined separately for charge and spin degrees of freedom, and employ it to recognize the decoupling of charge and spin degrees of freedom accompanying a~finite-system version of the Mott transition. 

The paper is organized as follows. In Sec.\ II we present the system Hamiltonian and summarize the EDABI method. In Sec.\ III the findings of Ref.\ \cite{Ryc04}, concerning the finite-size scaling of charge-energy gap, are revisited after taking larger systems under consideration. In Sec.\ IV we determine the local and pairwise entanglement, the latter for charge and spin degrees of freedom, and discuss their evolution with the interatomic distance and the system size. The conclusions are given in Sec.\ V.

\section{The exact diagonalization -- ab initio method (EDABI)}
The EDABI method, together with its application to the system of Fig.\ \ref{fig:chain}, have been presented in several works \cite{Ryc04,Kad14,Kad15,Spa01}. Here we briefly recall the main findings, to which we refer in the remaining parts of the paper. 

\subsection{The Hamiltonian for a~linear chain}
The analysis starts from the second-quantized Hamiltonian, which can be written in the form
\begin{multline}
  \label{hamchain}
%  H_{\rm chain}=
  \hat{\cal H}(\alpha,R)=
  \epsilon_a\sum_j\hat{n}_j 
  +\sum_{i<j,\sigma}t_{ij}\left(\hat{c}_{i\sigma}^{\dagger}\hat{c}_{j\sigma}+
  \hat{c}_{j\sigma}^{\dagger}\hat{c}_{i\sigma}\right) \\
  + U\sum_i\hat{n}_{i\uparrow}\hat{n}_{i\downarrow} 
  + \sum_{i<j}K_{ij}{\hat{n}_i}{\hat{n}_j} +V_{\rm ion-ion},
\end{multline}
where $\epsilon_a$ is the atomic energy (same for all $N$ sites upon applying periodic boundary conditions), $t_{ij}$ is the hopping integral between $i$-th and $j$-th site (we further set $t_{ij}\equiv{}-t$ if $i$ and $j$ are nearest neighbors, otherwise $t_{ij}\approx{}0$), $U$ is the intrasite Coulomb repulsion, $K_{ij}$ is the intersite Coulomb repulsion, and $V_{\rm ion-ion}=\sum_{i<j}e^2/R_{ij}$ (with the distance $R_{ij}\equiv{}R\min(|i-j|,N-|i-j|)$) expresses the repulsion of infinite-mass ions. 

The single-particle and  interaction parameters of the Hamiltonian $\hat{\cal H}(\alpha,R)$ (\ref{hamchain}), also marked in Fig.\ \ref{fig:chain}, can be defined as follows
\begin{align}
  \langle{}w_i|T|w_i\rangle &= 
  \int{}d^3r{}\,w_i^\star({\bf r})T({\bf r})w_j({\bf r}) \nonumber\\
  & \equiv{} \epsilon_a\delta_{ij}+t_{ij}(1-\delta_{ij}), 
  \label{epstijdef}
\end{align}
\begin{align}
  \langle{}w_iw_j|V|w_iw_j\rangle &= 
  \int{}d^3r{}d^3r'{}\,|w({\bf r})|^2V({\bf r}-{\bf r}')|w({\bf r}')|^2
  \nonumber\\
  & \equiv{} U\delta_{ij}+K_{ij}(1-\delta_{ij}),
  \label{ukijdef}
\end{align}
where $T$ is the~single particle Hamiltonian describing an electron in the medium of periodically arranged ions, and $V$ is the Coulomb repulsive interaction of two electrons. 
The Wannier functions are defined via atomic (Slater-type) functions, namely
\begin{equation}
  \label{wannfun}
  w_i({\bf r})=\sum_j\beta_{ij}\psi_j({\bf r}),
\end{equation}
where the Slater 1$s$ function $\psi_i({\bf r})=(\alpha^3/\pi)^{1/2}\exp(-\alpha{}|{\bf r}\!-\!{\bf R}_i|)$, with $\alpha$ being the inverse orbital size, here taken as variational parameter, and ${\bf R}_i$ being the position of $i$-th ion. The coefficients $\beta_{ij}$ in Eq.\ (\ref{wannfun}) can be uniquely defined by imposing that (i) $\langle{}w_i|w_k\rangle=\delta_{ik}$ and that (ii) $\langle{}w_i|\psi_i\rangle$ is maximal \cite{Ryc04,Rycphd}. 

The range of Coulomb interactions in the Hamiltonian  $\hat{\cal H}(\alpha,R)$ (\ref{hamchain}), quantified by parameters $K_{ij}\approx{}e^2/R_{ij}$ for $R_{ij}\gg{}R$, is {\em a~priori} limited only by the chain length $L\equiv{}NR$. 
Previous studies on linear chains, both {\em ab initio} ones \cite{Ste11,Hach06} and these starting from second-quantized model Hamiltonians \cite{Hoh12,Eji05}, usually have imposed some form of charge screening reducing the range of such interactions (leaving the on-site Coulomb repulsion only in the extreme case \cite{Sch90}). This can be partly justified by possible influence of the enviroment in condensed-matter realizations such as quantum wires \cite{Mey09} or self-organised chains \cite{Blu11}. A~slightly different situation occurrs for cold atom systems, where long-range dipole-dipole interactions may be relevant \cite{Mos15}. Apart from these experiment-related premises, long-range interactions usually lead to a~noticeable slowdown of the convergence for numerical methods such as DMRG or QMC \cite{Cap00,Fan99}. This is not the case for the EDABI method, as the determination of parameters $K_{ij}$ corresponds to a~relatively small portion of the overall computation time \cite{Bib15}, and the convergence of subsequent diagonalization in the Fock space (see the next subsection) is unaffected by the fact whether or not long-range interactions are included. Also, as charge screening becomes systematically less effective when the system dimensionality is reduced \cite{Kon77}, one can expect it to be insignificant in a~hypotetical realization of chains containing $N\leqslant{}20$ atoms.

\subsection{Single-particle basis optimization}
Next, each Slater function is approximated as follows
\begin{equation}
  \psi_i({\bf r}) \approx{} \alpha^{3/2}
  \sum_{q=1}^p{}B_q\left(\frac{2\Gamma_q^2}{\pi}\right)^{3/4}
%  \exp\left(-\alpha^2\Gamma_q^2|{\bf r}-{\bf R}_i|^2\right), 
  e^{-\alpha^2\Gamma_q^2|{\bf r}-{\bf R}_i|^2}, 
\end{equation}
where $p$ is the number of Gaussian functions truncating $\psi_j$, and $\{B_q,\Gamma_q\}$, $q=1,\dots,p$, are adjustable parameters chosen to minimize the atomic energy for $\alpha=1$ and a~given value of $p$. Here we set $p=3$, for which the deviation from the exact energy for $1s$ function is lower that $1\%$ \cite{Rycappa}. Subsequently, the parameters $\epsilon_a$, $t_{ij}$, $U$, and $K_{ij}$ are calculated from Eqs.\ (\ref{epstijdef}), (\ref{ukijdef}) as functions of $\alpha$ and $R$. The Hamiltonian $\hat{\cal H}(\alpha,R)$ (\ref{hamchain}) is diagonalized numerically in the Fock space, using the Lanczos algorithm, and the orbital size is optimized to find $\alpha=\alpha_{\rm min}$ corresponding to the minimal GS energy $E_G(N)$ for each $R$. Defining the efective atomic energy as
\begin{equation}
  \label{eaeffdef}
  \epsilon_a^{\rm eff}=\epsilon_a
  +\frac{1}{N}\bigg(\sum_{i<j}K_{ij}+V_{\rm ion-ion}\bigg), 
\end{equation}
one finds that $\alpha_{\rm min}$ and other parameters converges rapidly with $N$. This observation allows us to speed up computations by using the values obtained for smaller systems ($N\geqslant{}10$) to perform extrapolation with $1/N\rightarrow{}0$. (The results are listed in Table~\ref{tab:param}.) 

\begin{table*}
\caption{ \label{tab:param}
  Microscopic parameters (specified in eV -- unless stated otherwise) of the Hamiltonian  $\hat{\cal H}(\alpha,R)$ (\ref{hamchain}) with $\alpha=\alpha_{\rm min}$ minimizing the ground-state (GS) energy $E_G$. The Bohr radius is $a_0=0.529\,$\AA. The effective atomic level $\epsilon_a^{\rm eff}$ is defined by Eq.\ (\ref{eaeffdef}); the remaining symbols are $t\equiv{}-t_{ij}$ for $j=i\pm{}1$ (mod $N$), $K_m\equiv{}K_{|i-j|}$ (with $m=1,2,3$), and  the correlated hopping integral $V\equiv\langle{}w_iw_i|V|w_iw_{i\pm{}1}\rangle$, quantifying the largest neglected term in  $\hat{\cal H}(\alpha,R)$. The numerical extrapolation with $1/N\rightarrow{}0$ is performed for all parameters. 
}
\begin{tabular}{c|c|c|c|c|c|c|c|c|c}
\hline\hline
$\ R/a_0\ $  &  $\ \alpha_{\rm min}a_0\ $  &  $\ \epsilon_a^{\rm eff}\ $  &  $\ t\ $  &  $\ U\ $  &  $\ K_1\ $  &  $\ K_2\ $  &  $\ K_3\ $  & $\ V\ $  &  $\ E_G/N\ $  \\ \hline
1.5  &  1.363  &  $\ \ \ $1.36$\ $  &  $\ $11.31$\ $  &  $\ $27.95$\ $  &  $\ $15.85$\ $  &  $\ $9.08$\ $  &  $\ $6.08$\ $  & $\ $-0.597$\ $ & $\ $-10.19$\ $  \\ 
2.0  &  1.220  &  $\ $-7.48  &  $\ \ $6.02$\ $  &  23.58  &  12.40  &  6.82  &  4.54  & -0.324 &  -12.65  \\ 
2.5  &  1.122  &  10.85  &  $\ \ $3.60$\ $  &  20.83  &  10.20  &  5.46  &  3.63  & -0.203 &  -13.32  \\ 
3.0  &  1.062  &  12.27  &  $\ \ $2.32$\ $  &  19.14  &  $\ $8.69  &  4.54  &  3.02  & -0.150 &  -13.48  \\ 
3.5  &  1.031  &  12.90  &  $\ \ $1.57$\ $  &  18.16  &  $\ $7.58  &  3.89  &  2.60  & -0.128 &  -13.50  \\ 
4.0  &  1.013  &  13.20  &  $\ \ $1.08$\ $  &  17.57  &  $\ $6.71  &  3.40  &  2.27  & -0.119 &  -13.50  \\ 
5.0  &  1.004  &  13.43  &  $\ \ $0.51$\ $  &  17.12  &  $\ $5.43  &  2.72  &  1.81  & -0.096 &  -13.50  \\ 
6.0  &  1.001  &  13.48  &  $\ \ $0.23$\ $  &  16.99  &  $\ $4.53  &  2.27  &  1.51  & -0.058 &  -13.50  \\ 
7.0  &  1.000  &  13.49  &  $\ \ $0.10$\ $  &  16.97  &  $\ $3.89  &  1.99  &  1.29  & -0.027 &  -13.50  \\ 
\hline\hline
\end{tabular}
\end{table*}

\section{Finite-size scaling for the charge-energy gap}

\begin{figure}
\centerline{\includegraphics[width=0.9\linewidth]{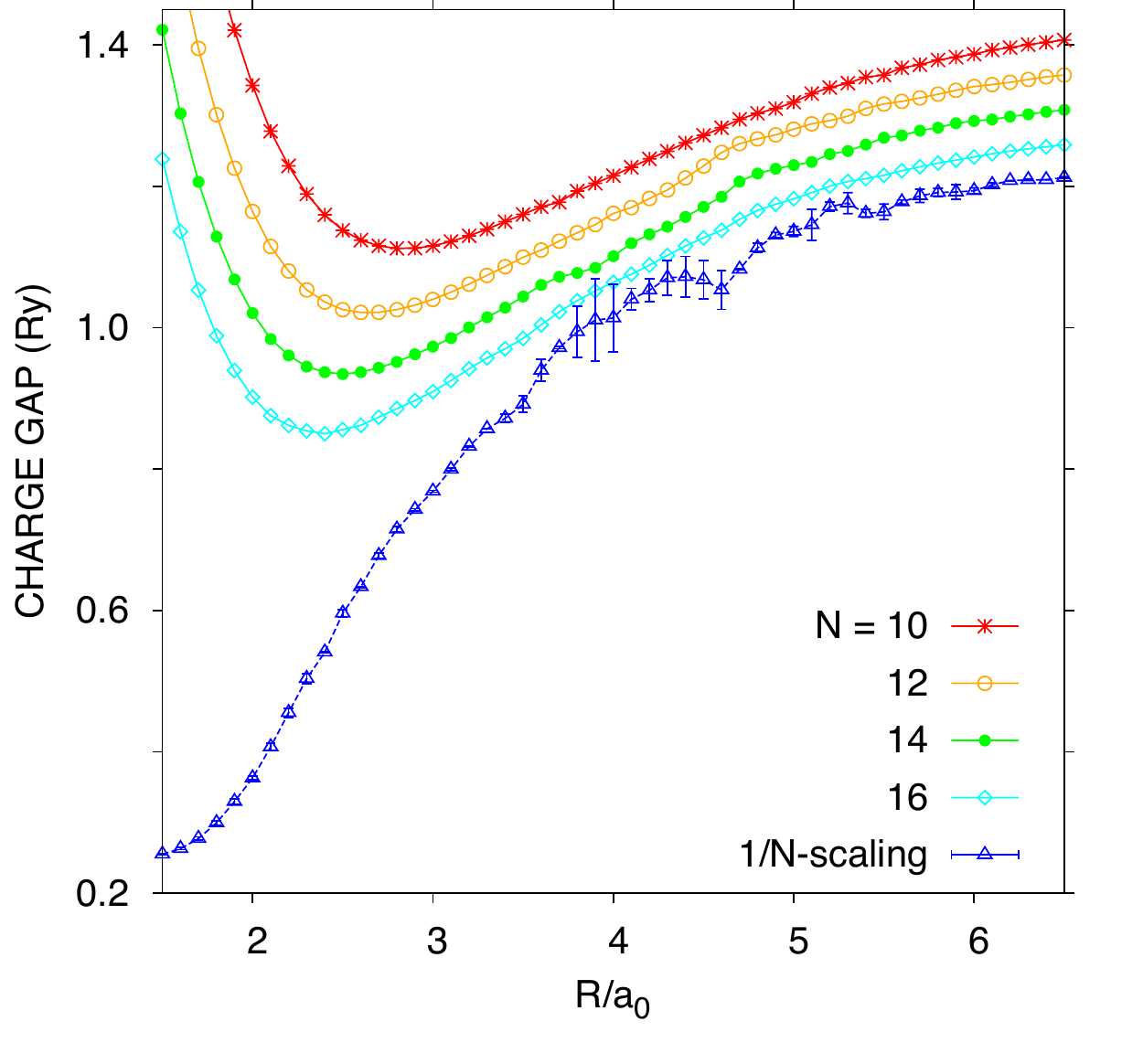}}
\caption{ \label{fig:delc-half}
  Charge-energy gap as a~function of the interatomic distance $R/a_0$ 
  (with the Bohr radius $a_0=0.529\,$\AA) for chains of $N=10-16$ atoms 
  at the {\em half filling} ($N_{\rm el}=N$). 
  Datapoints were shifted vertically by $0.05\,$Ry for $N=16$ (diamonds), 
  $0.10\,$Ry for $N=14$ (solid circles), $0.15\,$Ry for $N=12$ (open circles), 
  or $0.20\,$Ry for $N=10$ (stars). 
  The results of the finite-size scaling with $1/N\rightarrow{}0$, 
  obtained via the Richardson extrapolation of the second order \cite{Bur10},  
  are also shown (triangles).  Lines are guides for the eye only.
}
\end{figure}

\begin{figure}
%\centerline{\includegraphics[width=0.9\linewidth]{delc-quart}}
\centerline{\includegraphics[width=0.9\linewidth]{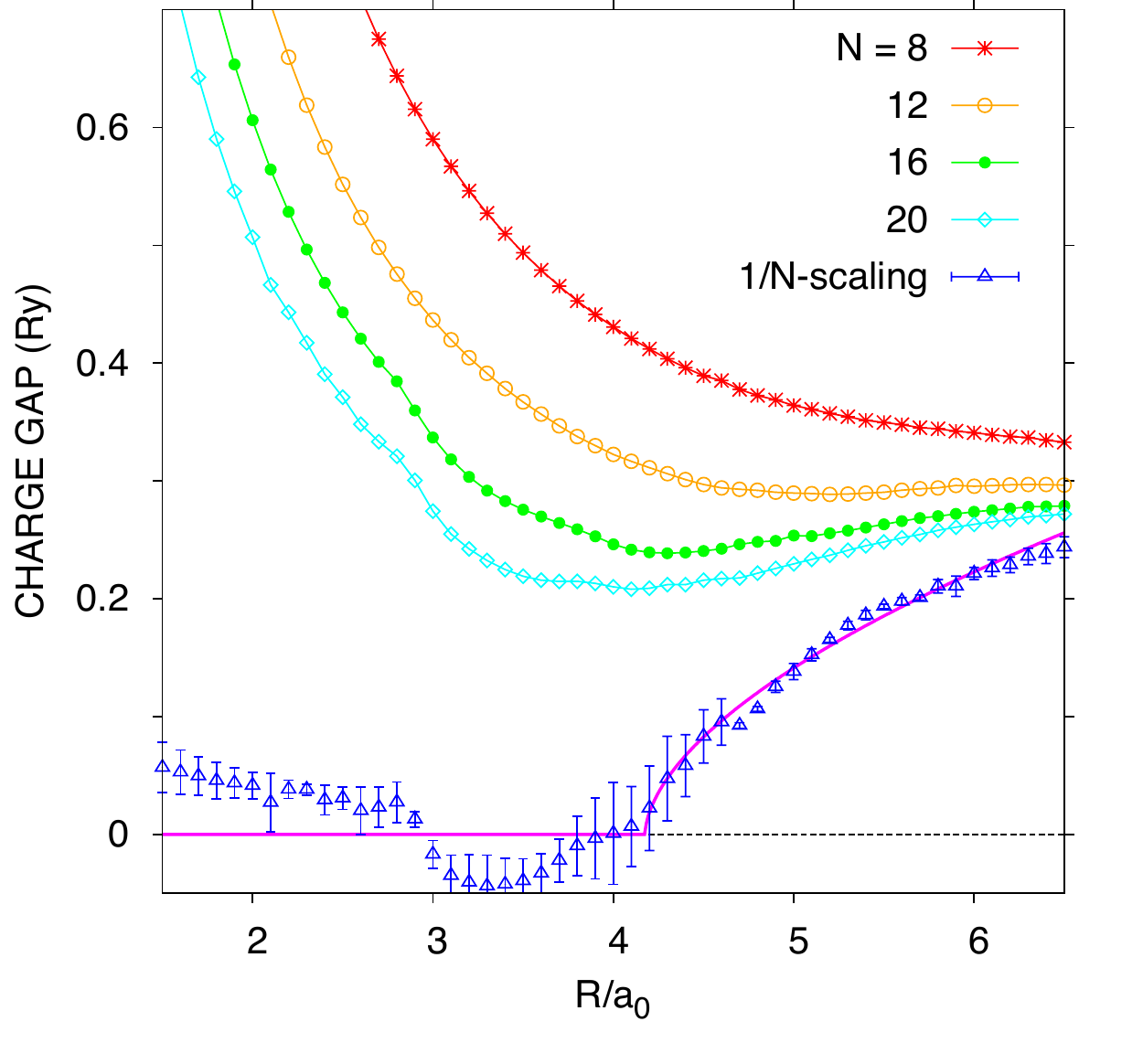}}
\caption{ \label{fig:delc-quart}
  Same as Fig.\ \ref{fig:delc-half} but for chains of $N=8-20$ atoms
  at the {\em quarter filling} ($N_{\rm el}=N/2$). No artificial datashifts 
  were applied this time. Horizontal dashed line marks $\Delta{}E_C=0$, 
  solid purple line represents the best-fitted function given by 
  Eqs.\ (\ref{empdelc}) and (\ref{empdpar}), 
  indicating the Mott metal-insulator transition in the 
  $1/N\rightarrow{}0$ limit, occurring for $R=R_c\approx{}4.2\,a_0$. 
}
\end{figure}

A~standard numerical approach \cite{Hir85,Hoh12}, allowing one to determine whether a~correlated system described by the Hamiltonian such as given by Eq.\ (\ref{hamchain}) is metallic or insulating in the ground state, involves calculating the charge-energy gap according to 
\begin{equation}
  \label{delcdef}
  \Delta{}E_C^{N_{\rm el}}(N)= 
  E_G^{N_{\rm el}+1}(N)+E_G^{N_{\rm el}-1}(N)-2E_G^{N_{\rm el}}(N),
%  \Delta{}E_C(N,N_{\rm el})= 
%  E_G(N,N_{\rm el}+1)+E_G(N,N_{\rm el}-1)-2E_G(N,N_{\rm el}),
\end{equation}
where $N_{\rm el}$ is the number of electrons. Next, the limit of $1/N\rightarrow{}0$ (with $N_{\rm el}/N=\,$const) is to be taken numerically. The limiting value of $\Delta{}E_C\rightarrow{}0$ indicates the metallic phase, whereas a~nonzero value indicates the insulating phase \cite{spingapfoo}. In the numerical examples presented here and in Sec.\ IV we consider two different physical situations: the {\em half filling} $N_{\rm el}=N$, with $N\leqslant{}16$, and the {\em quarter filling} $N_{\rm el}=N/2$, with $N\leqslant{}20$; we further restrict ourselves to even values of $N$ and $N_{\rm el}$. Due to the total spin conservation it is sufficient, for each pair $(N,N_{\rm el})$, to look for the ground-state energy $E_G^{N_{\rm el}}(N)$ in Eq.\ (\ref{delcdef}) in the subspace characterized by the total $z$-th component of spin $S_z=0$. Analogously, the values $E_G^{N_{\rm el}\pm{}1}(N)$  can be found by choosing $S_z=1/2$. In particular, the largest considered subspace dimension corresponds to $N=20$, $N_{\rm el}\!+\!1=11$, and is equal to $600\,935\,040$. Moreover, we impose the periodic ($t_{i,j+N}=t_{ij}$) or antiperiodic ($t_{i,j+N}=-t_{ij}$) boundary conditions to minimize $E_G^{N_{\rm el}}(N)$ for $N_{\rm el}=4k+2$ or $N_{\rm el}=4k$, respectively, with $k$-integer (see Table~\ref{tab:rstara}).

\begin{table}
\caption{ \label{tab:rstara}
  Boundary conditions BC (with $+/-$ marking the periodic/antiperiodic BC) 
  and dimension of the largest subblock of the Hilbert space (with the total
  $z$-th component of spin $S_z=0$) for chains studied in the paper at 
  the quarter filling (i.e., $N_{\rm el}=N/2$).  The last column specifies
  the values of the interatomic distance $R_\star$ at which the spin 
  concurrence vanishes, with the interpolation errorbars for the last digit 
  given in parenthesis. (See Sec.\ IV for details.) 
}
\begin{tabular}{c|c|c|r|c}
\hline\hline
  $\,\ N\,\ $ & $\ N_{\rm el}\ $ & $\ $BC$\ $
  & $\ $dim$\,{H}(S_z^{\rm tot}\!\!=\!0)\ $ & $R_\star/a_0$ \\ 
\hline
%%
%% ==>  STARE "R_\star" <==  
%% $\ $8  & $\ $4  &  $-$ & 784$\ $ & $\,\ $5.59(3)$\,\ $ \\
%%    12  & $\ $6  &  $+$ & 48,400$\ $ &   4.17(1)  \\
%%    16  & $\ $8  &  $-$ & 3,312,400$\ $ &   3.82(1)  \\
%%    20  &    10  &  $+$ & $\ $240,374,016$\ $ &   3.53(3)  \\
%%
 $\ $8  & $\ $4  &  $-$ & 784$\ $ & $\,\ $6.53(3)$\,\ $ \\
    12  & $\ $6  &  $+$ & 48,400$\ $ &   5.22(2)  \\
    16  & $\ $8  &  $-$ & 3,312,400$\ $ &   4.38(1)  \\
    20  &    10  &  $+$ & $\ $240,374,016$\ $ &   4.51(1)  \\
\hline\hline
\end{tabular}
\end{table}

The numerical results, corresponding to the Hamiltonian  $\hat{\cal H}(\alpha,R)$ (\ref{hamchain}) with the microscopic parameters listed in Table~\ref{tab:param}, are presented in Figs.\ \ref{fig:delc-half} and \ref{fig:delc-quart}. We notice that the datasets for $N\leqslant{}14$ at the half filling as well as for $N\leqslant{}16$ at the quarter filling were presented in Ref.\ \cite{Ryc04}, were we also used the Richardson extrapolation of the second order \cite{Bur10} to perform finite-size scaling with $1/N\rightarrow{}0$ for each value of $R$.  After adding the new datasets for $N=16$, $N_{\rm el}=N$ and $N=20$, $N_{\rm el}=N/2$ (both depicted with open diamonds) the earlier conclusions regarding the system ground state are further supported: Namely, the extrapolation with $1/N\rightarrow{}0$ (open triangles)  at $N_{\rm el}=N$ leads to $\Delta{}E_C>0$ for any accessible value of $R/a_0$ indicating the insulating phase (see Fig.\ \ref{fig:delc-half}). At $N_{\rm el}=N/2$, the finite-size scaling results can be rationalized with the empirical function
\begin{equation}
  \label{empdelc}
  \Delta{}E_C^{N/2}(1/N\!\rightarrow{}\!0)\approx 
  \begin{cases}
    0 & \ \text{if } R<R_c, \\
    \Delta_0\left[(R\!-\!R_c)/a_0\right]^{\gamma}
    & \ \text{if } R\geqslant{}R_c, \\
  \end{cases}
\end{equation}
with the best-fitted parameters (numbers in parentheses are standard deviations for the last digit)
\begin{gather}
  R_{\rm c}= 4.18(2)\,a_0, \ \ \ \ \Delta_0=0.158(3)\,\text{Ry},\nonumber \\
  \gamma= 0.57(3),  \label{empdpar}
\end{gather}
depicted with solid purple line in Fig.\ \ref{fig:delc-quart}. 
The gap opening at $R=R_{\rm c}$ indicates the Mott transition. 

For a~sake of conciseness, our discussion of finite-size scaling estimates of the Mott transition is limited to the charge-energy gap. A~more detailed analysis, presented in Refs.\ \cite{Ryc04,Spa01,Rycphd} and involving calculations of the electron momentum distribution, the Drude weight, as well as the so-called modern theory of polarization \cite{Res02}, justify the transition appearance in the $N_{\rm el}=N/2$ case, which coincides with the results for related parametrized model studies \cite{Hoh12,Eji05,Sch90,Mos15}. In the $N_{\rm el}=N$ case, the situation is of a~slightly more complex nature: Apart from a~nonzero gap for any $R$, following from the finite-size scaling, several GS and dynamical characteristics (in particular -- the Drude weight, see Ref.\ \cite{Ryc04}) exhibit, for any finite $N$, a~crossover behavior between a~partly localized quantum liquid, appearing for small $R$, and a~fully-reconstructed Mott insulator, typically appearing for 
$R/a_0\gtrsim{}4$. These findings coincides with more recent variational Monte-Carlo studies of hydrogenic chains (see Ref.\ \cite{Ste11}), and can be attributed to the increasing (with growing $R$) role of electron correlations in either the exact or variational GS wavefunction.

It is worth to mention here that some surprisingly efficient variational methods, such as presented in Ref.\ \cite{Ste11}, become significantly less efficient when treating open-shell configurations in attempt to calculate $\Delta{}E_C$ from Eq.\ (\ref{delcdef}), or generally when utilized away from the half filling ($N_{\rm el}\neq{}N$). Also for this reason, GS correlation functions signalling a~finite-system version of the Mott transition for $N$ small enough to be treated with some more flexible numerical techniques, are desired.

\section{Reduced density matrix and quantum entanglement}
In this section we derive explicit representations of the reduced density operator $\hat{\rho}_A=\mbox{Tr}_A|\Psi\rangle\!\langle\Psi|$ (where Tr$_A$ stands for tracing over all degrees of freedom except from these characterizing a~selected subsystem $A$) relevant when discussing the local entanglement, the pairwise entanglement for charge degrees of freedom, and the pairwise entanglement for spin degrees of freedom. The derivations remain valid for a~generic spin--$1/2$ fermionic system at any pure state $|\Psi\rangle$, which can be represented assuming four basis states for each site $j$, namely: 
\begin{equation}
\label{basenjud}
\left\{\,|n_{j\uparrow}n_{j\downarrow}\rangle\,\right\}\equiv\left\{\,
|0\rangle_j,  |\!\uparrow\rangle_j, |\!\downarrow\rangle_j,  |\!\uparrow\downarrow\rangle_j\,\right\}; 
\end{equation}
although the numerical examples, considered in search for entanglement-based estimates of the Mott transition, all correspond to the ground state of a~linear chain described by the Hamiltonian $\hat{\cal H}(\alpha,R)$ (\ref{hamchain}) with the microscopic parameters taken from Table~\ref{tab:param}.

%\subsection{Preliminaries}
\subsection{General considerations}
Let us consider a~general example at first: the quantum system, which can be divided into two distinct subsystems $A$ and $B$. A~pure state $|\Psi\rangle$ can be represented as follows 
\begin{equation}
\label{psialpbet}
  |\Psi\rangle=\sum_{\alpha\beta}\Psi_{\alpha\beta}|\alpha\rangle\otimes|\beta\rangle,
\end{equation}
where $\Psi_{\alpha\beta}$ denotes a~complex probability amplitude corresponding to a~basis state of the full system $|\alpha\rangle\otimes|\beta\rangle$, while $\{|\alpha\rangle\}$ and $\{|\beta\rangle\}$ are complete basis sets for the subsystems $A$ and $B$ (respectively). The reduced density operator $\hat{\rho}_A$
is defined as:
\begin{align}
  \hat{\rho}_A&=
  \mbox{Tr}_A|\Psi\rangle\!\langle\Psi|\equiv
  \sum_\beta\langle\beta|\Psi\rangle\!\langle\Psi|\beta\rangle \nonumber \\
  &=
  \sum_{\beta,\alpha,\alpha'}\Psi_{\alpha\beta}{\Psi}_{\alpha'\beta}^{\star}
  |\alpha\rangle\!\langle\alpha'|,
\end{align}
where the last equality follows from Eq.\ (\ref{psialpbet}) and $\Psi_{\alpha\beta}^{\star}$ denotes the complex conjugate of $\Psi_{\alpha\beta}$. Subsequently, the matrix elements of $\hat{\rho}_A$ are given by
\begin{equation}
\label{rhoma}
  \rho_{\alpha,\alpha'}\equiv\langle{\alpha}|\hat{\rho}_A|{\alpha'}\rangle=
  \sum_\beta\Psi_{\alpha\beta}{\Psi}_{\alpha',\beta}^{\star}.
\end{equation}

We define now projection operators $P_\alpha$ and $P_\beta$, associated with 
the basis states $|\alpha\rangle$ and $|\beta\rangle$, via 
\begin{equation}
\label{palpdef}
  P_{\alpha}|\alpha'\rangle=\delta_{\alpha\alpha'}|\alpha\rangle, \ \ \ \ 
  P_{\beta}|\beta'\rangle=\delta_{\beta\beta'}|\beta\rangle.
\end{equation}
Useful properties, directly following from Eq.\ (\ref{palpdef}) are
\begin{gather}
P_\alpha^2=P_\alpha, \ \ \ \  P_\beta^2=P_\beta, \label{palpsqu} \\
  \sum_\alpha P_\alpha = \sum_\beta P_\beta = 1, \label{palpsum} 
\end{gather}
where Eq.\ (\ref{palpsum}) also employs the completeness of the basis sets $\{|\alpha\rangle\}$ and $\{|\beta\rangle\}$.  
Next, we define the transfer operator 
\begin{equation}
\label{tradef}
T_{\alpha\alpha'}|{\alpha'}\rangle=|{\alpha}\rangle, \ \ \ \ 
\langle{\alpha'}|T_{\alpha\alpha'}^{\dagger}=\langle{\alpha}|, 
\end{equation}
which is unitary, i.e.\ 
\begin{equation}
\label{trainv}
T_{\alpha\alpha'}^{\dagger}=(T_{\alpha\alpha'})^{-1}\equiv 
T_{\alpha'\alpha}. 
\end{equation}
Explicit forms of  $P_\alpha$, $P_\beta$, and $T_{\alpha\alpha'}$, corresponding to the particular splittings of the system into $A$ and $B$ subsystems, are to be specified later in terms of the operators $\{\hat{c}_{i\sigma},\hat{c}_{i\sigma}^{\dagger}\}$ for lattice spin--$1/2$ fermions. Here we notice that as the indices $(\alpha,\alpha')$ refer the subsystem $A$, whereas the index $\beta$ refers to the subsystem $B$, we have
\begin{equation}
\label{ptcomm}
\left[P_\alpha,P_\beta\right]=0, \ \ \ \ 
\left[P_\beta,T_{\alpha\alpha'}\right]=0,
\end{equation}
for all  $(\alpha,\alpha')$ and $\beta$. 

With the help of operators $P_{\alpha}$ and $P_{\beta}$ (\ref{palpdef}) one can write down, for $|\Psi\rangle$ represented according to Eq.\ (\ref{psialpbet}),
\begin{align}
  P_\alpha P_\beta|\Psi\rangle &= \Psi_{\alpha\beta}\,
  |\alpha\rangle\otimes|\beta\rangle, \\
  \langle\Psi|P_{\alpha'}P_\beta &= \langle\alpha'|\otimes\langle\beta|
  \,{\Psi}_{\alpha'\beta}^{\star}. 
\end{align}
In turn, the reduced density matrix, as given by the rightmost equality in Eq.\ (\ref{rhoma}), can be rewritten as 
\begin{align}
  \rho_{\alpha\alpha'}
  &=\sum_{\beta}\langle\Psi|P_{\alpha'}P_{\beta}
  T_{\alpha'\alpha}P_\alpha P_\beta|\Psi\rangle \nonumber\\
  &=
  \langle\Psi|P_{\alpha'}T_{\alpha'\alpha}P_\alpha|\Psi\rangle, 
\end{align}
with the last equality following from Eqs.\ (\ref{palpsqu}), (\ref{palpsum}), and (\ref{ptcomm}). Finally, using Eq.\ (\ref{trainv}), we find
\begin{equation}
\label{rhostar}
  \left(\rho_{\alpha\alpha'}\right)^{\star} = 
  \langle{\Psi}|P_{\alpha}T_{\alpha\alpha'}P_{\alpha'}|{\Psi}\rangle.
\end{equation}
Remarkably, the reduced density matrix elements in Eq.\ (\ref{rhostar}) are expressed by pure-state expectation values (correlation functions) of the operators acting only on the subsystem $A$. This feature is further explored in the remaining parts of this section.

\subsection{Entanglement entropy}
The local entanglement \cite{Zan02} exhibits quantum correlations between the local state of a~selected $j$-th site (subsystem $A$) and the rest of the system ($B$). As the basis set $\{|\alpha\rangle\}$ is simply given by Eq.\ (\ref{basenjud}), one can write down the corresponding projection operators
\begin{multline}
\label{palploc}
  \left(P_\alpha\right) =
  \left[\, (1-n_\uparrow)(1-n_\downarrow),\ n_\uparrow (1-n_\downarrow),\ \right. \\
  \left.(1-n_\uparrow)n_\downarrow,\ n_\uparrow n_\downarrow \,\right]^T,
\end{multline}
and the transfer operators 
\begin{equation}
\label{traloc}
  \left(T_{\alpha\alpha'}\right)=\left(
  \begin{array}{cccc}
  1 & c_\uparrow & c_\downarrow & c_\downarrow c_\uparrow \\
  c_\uparrow^\dagger & 1 & c_\uparrow^\dagger c_\downarrow & -c_\downarrow \\
  c_\downarrow^\dagger & c_\downarrow^\dagger c_\uparrow & 1 & c_\uparrow \\
  c_\uparrow^\dagger c_\downarrow^\dagger & -c_\downarrow^\dagger & c_\uparrow^\dagger & 1 \\
  \end{array}\right),
\end{equation}
where we have omitted the site index $j$ for brevity. 
For the system with the total spin and charge conservation, such as our linear chain [see Eq.\ (\ref{hamchain})], we have 
\begin{gather}
\langle{c_\sigma}\rangle=
\langle{c_\sigma^\dagger}\rangle=0, 
\ \ \ \ 
\langle{ c_\sigma c_{\bar{\sigma}} }\rangle=
\langle{c_\sigma^\dagger c_{\bar{\sigma}}^\dagger}\rangle=0, 
\nonumber \\
\text{and}
\ \ \ \ 
\langle{c_\sigma^\dagger c_{\bar{\sigma}}}\rangle=
\langle{c_{\bar{\sigma}}^\dagger c_\sigma}\rangle=0, \label{cavezero}
\end{gather}
(with $\overline{\sigma}$ being the spin index opposite to $\sigma$) providing that the averaging takes place over an eigenstate $|{\Psi}\rangle$ of the system Hamiltonian (not necessarily the ground state). Substituting the expressions for $P_\alpha$ (\ref{palploc}) and $T_{\alpha\alpha'}$ (\ref{traloc}) into Eq.\ (\ref{rhostar}) and taking Eqs.\ (\ref{cavezero}) into account we get, after a straightforward algebra, 
\begin{equation}
  \label{rholoc}
  \left(\rho_{\alpha\alpha'}\right)=
  \mbox{diag}\left( u_+,w_1,w_2,u_- \right)
\end{equation}
with
\begin{gather}
  u_+ = \left<{(1-n_{j\uparrow})(1-n_{j\downarrow})}\right>,\ \ \ \ 
  w_1 = \left<{n_{j\uparrow}(1-n_{j\downarrow})}\right>,
\nonumber \\
  w_2 = \left<{(1-n_{j\uparrow})n_{j\downarrow}}\right>,\ \ \ \ 
  u_- = \left<{n_{j\uparrow}n_{j\downarrow}}\right>. \label{locelem}
\end{gather}
Eq.\ (\ref{rholoc}) corresponds to the reduced density operator of the well-known form
\begin{multline}
\label{rholocop}
  \hat{\rho}_A = 
  \sum_{\alpha\alpha'}\rho_{\alpha\alpha'}|\alpha\rangle\!\langle\alpha'|= \\
  u_+|0\rangle\!\langle{}0| +
  w_1|\!\uparrow\rangle\!\langle{}\!\uparrow\!| +
  w_2|\!\downarrow\rangle\!\langle{}\!\downarrow\!| +
  u_-|\!\uparrow\downarrow\rangle\!\langle{}\!\uparrow\downarrow\!|, 
\end{multline}
where we have used the basis given by Eq.\ (\ref{basenjud}). 

A~quantitative measure of the entanglement between the state of $j$-th site and that of the remaining $N-1$ sites is given by the von Neumann entropy 
\begin{equation}
\label{evdef}
  E_v = -u_+\log_2u_+ - w_1\log_2w_1 - w_2\log_2w_2 -u_-\log_2u_-.
\end{equation}
If the translational invariance of the system is imposed, one can define the particle density $n$ and the average number of double occupancies $d$, following
\begin{equation}
  \langle{}n_{j\uparrow}\rangle=\langle{}n_{j\uparrow}\rangle=\frac{N_{\rm el}}{2N}
  \equiv{}\frac{n}{2}, 
  \ \ \ \ \ \ 
  \langle{}n_{j\uparrow}n_{j\uparrow}\rangle\equiv{}d,
\end{equation}
where we have further assumed that the averaging takes place over an eigenstate characterized by the total $z$-th component of spin $S_z=0$. Subsequently, Eq.\ (\ref{locelem}) can be rewritten as 
\begin{equation}
\label{loceltinv}
  u_+=1-n+d, \ \ \ \ w_1=w_2=\frac{n}{2}-d, \ \ \ \ 
  u_-=d. 
\end{equation}

In particular, for the ground state of the Hamiltonian $\hat{\cal H}(\alpha,R)$ (\ref{hamchain}), for which the effective interaction between electrons can roughly be estimated by $\sim{}U-K_1$ and is always repulsive, one can show that
\begin{equation}
\label{dbounds}
  \frac{|n-1|+n-1}{2}\leqslant{}d\leqslant{}\frac{n^2}{4} \ \ \ \ 
  \text{for}\ \ \ \ 0\leqslant{}n\leqslant{}2,
\end{equation}
where the lower bound for $d$ corresponds to the so-called {\em strong-correlations limit} approached for $R\gg{}a_0$ (as the bandwidth $W\equiv{}4t\ll{}U-K_1$, see Table~\ref{tab:param}), whereas the upper bound for $d$ corresponds to the {\em free-electrons limit} approached for $R\lesssim{}2a_0$ ($W\gg{}U-K_1$). As the charge fluctuations are given by $\mbox{Var}\{n_j\}=\langle{}n_j^2\rangle-\langle{}n_j\rangle^2\equiv{}n-n^2+2d$, the above-mentioned limits coincide (respectively) with the minimal and the maximal fluctuations for a~given $n$. Moreover, the bounds for $d$ in Eq.\ (\ref{dbounds}) can be mapped, via Eq.\ (\ref{loceltinv}), onto  
\begin{equation}
\label{evbounds}
E_v^{\rm corr}(n)\leqslant{}E_v\leqslant{}E_v^{\rm free}(n),
\end{equation}
with
\begin{multline}
  E_v^{\rm corr}(n)= -|1-n|\log_2|1-n| \\
  -\left(1-|1-n|\right)\log_2\left(\frac{1-|1-n|}{2}\right) \label{evboundc}
\end{multline}
for the strong-correlations limit, and
\begin{equation}
  E_v^{\rm free}(n)= 
  -n\log_2\frac{n}{2}-(2-n)\log_2\left(1-\frac{n}{2}\right) \label{evboundf}
\end{equation}
for the free-electrons limit. 
For instance, Eqs.\ (\ref{evbounds}), (\ref{evboundc}), and (\ref{evboundf}) lead to 
\begin{equation}
  \label{evbounds:half}
  1\leqslant{}E_v\leqslant{}2\ \ \ 
  \text{for}\ \ \ N_{\rm el}=N,
\end{equation}
and
\begin{equation}
  \label{evbounds:quart}
  \frac{3}{2}\leqslant{}E_v\leqslant{}4-\frac{3}{2}\log_2{}3
  \approx{}1.623\ \ \ 
  \text{for}\ \ \  N_{\rm el}=\frac{N}{2}.
\end{equation}
Remarkably, the strong-correlations limit corresponds to the minimal $E_v$, whereas the free-electrons limit corresponds to the maximal $E_v$ (for any $0\leqslant{}n\leqslant{}2$). This is because when the free-electrons limit is approached, ground-state wavefunction can be truncated by a~single Slater determinant composed of fully delocalized single-particle functions (Bloch functions), resulting in the maximal entanglement between $j$-th site and the remaining $N-1$ sites \cite{Ryc06d,Ste11}. Repulsive interactions between electrons lead to the correlation-induced suppression of $\mbox{Var}\{n_j\}$, and to the decreasing $E_v$ with growing the interaction-to-bandwidth ratio.

\begin{figure}
\centerline{\includegraphics[width=0.9\linewidth]{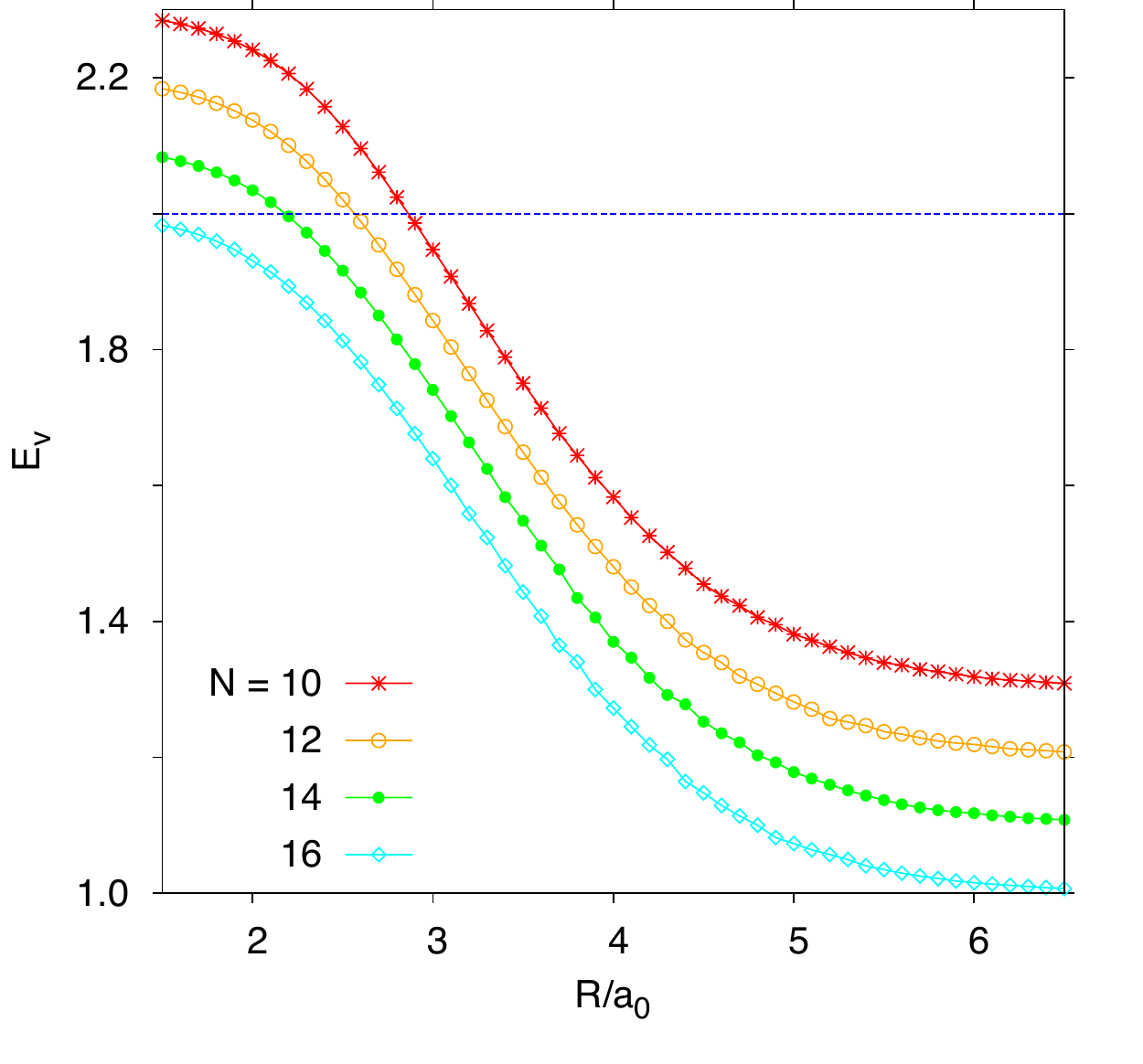}}
\caption{ \label{fig:evon-half}
  Local entanglement $E_v$ [see Eqs.\ (\ref{evdef})--(\ref{loceltinv})] 
  as a~function of the interatomic distance $R/a_0$ for $N=10-16$ 
  and $N_{\rm el}=N$. Datapoints for $N=16$ are shown unmodified, the others
  were shifted vertically by $0.1$ ($N=14$), $0.2$ ($N=12$), 
  or $0.3$ ($N=10$). Horizontal dashed line marks $E_v^{\rm free}(n)$
  given by Eq.\ (\ref{evboundf}); other lines are guides for the eye only.
}
\end{figure}

\begin{figure}
\centerline{\includegraphics[width=0.9\linewidth]{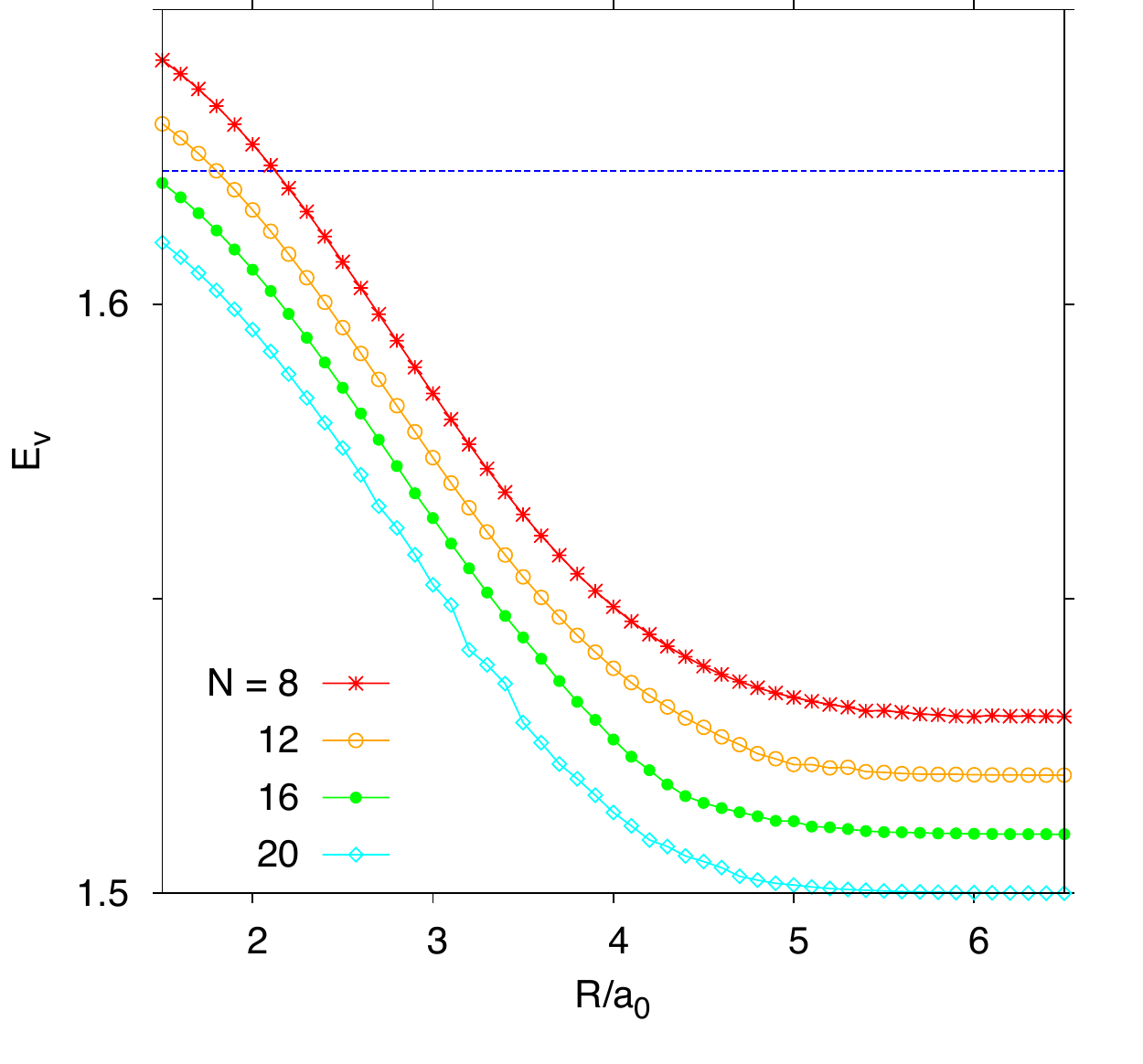}}
\caption{ \label{fig:evon-quart}
  Same as Fig.\ \ref{fig:evon-half}, but for $N=8-20$ and $N_{\rm el}=N/2$. 
  The vertical shifts are: $0$ ($N=20$), $0.01$ ($N=16$), $0.02$ ($N=12$), 
  and $0.03$ ($N=8$). 
}
\end{figure}

General findings, presented briefly in the above, are now illustrated with the numerical examples for  $N_{\rm el}=N$ and $N_{\rm el}=N/2$ (see  Figs.\ \ref{fig:evon-half} and \ref{fig:evon-quart}). In both cases, the values of $E_v$ are close to the upper bounds given by Eqs.\ (\ref{evbounds:half}) and (\ref{evbounds:quart}) for the smallest considered value of $R/a_0=1.5$, and systematically decrease with growing $R$,  gradually approaching the lower bounds in Eqs.\ (\ref{evbounds:half}) and (\ref{evbounds:quart}). Also, a remarkably fast convergence of $E_v$ with growing $N$ is observed for any $R$, making it necessary to apply vertical shifts to the datasets in Figs.\  \ref{fig:evon-half} and \ref{fig:evon-quart}.  
%% TUTAJ PISZEMY! %% 
We further notice that $E_v\approx{}3/2$ starting from $R\gtrsim{}5\,a_0$ for $N_{\rm el}=N/2$, whereas for $N_{\rm el}=N$ we still have $E_v>1$ in this range. However, the smooth evolution of $E_v$ with $R$ is observed for both $N_{\rm el}=N$ and $N_{\rm el}=N/2$, making it difficult to consider the local entanglement as an estimate of the Mott transition. Pairwise entanglement is discussed next as the other candidate.

\subsection{The fermionic concurrence}
We consider now the subsystem $A$ consists of two spatially-separate quantum bits ({\em qubits}), one of which is associated with $i$-th lattice site and the other with $j$-th site. Each individual qubit can be realized employing charge or spin degrees of freedom. 

For instance, if {\em charge qubits} are under consideration, one can choose the basis set for $A$ in a~fixed-spin sector $\sigma=\,\uparrow$, namely 
\begin{align}
  \left\{\,|\alpha\rangle\,\right\}^{c} =& \left\{\, 
    {|0\rangle}_i\otimes{|0\rangle}_j, 
    {|0\rangle}_i\otimes{|\!\uparrow\rangle}_j, \right. \nonumber \\  
    & \ \ \left. 
    {|\!\uparrow\rangle}_i\otimes{|0\rangle}_j, 
    {|\!\uparrow\rangle}_i\otimes{|\!\uparrow\rangle}_j 
  \,\right\}, 
\end{align}
whereas for {\em spin qubits} we have 
\begin{align}
  \left\{\,|\alpha\rangle\,\right\}^{s} =& \left\{\, 
    {|\!\uparrow\rangle}_i\otimes{|\!\uparrow\rangle}_j, 
    {|\!\uparrow\rangle}_i\otimes{|\!\downarrow\rangle}_j, \right. \nonumber \\  
    & \ \ \left. 
    {|\!\downarrow\rangle}_i\otimes{|\!\uparrow\rangle}_j, 
    {|\!\downarrow\rangle}_i\otimes{|\!\downarrow\rangle}_j 
  \,\right\}.  
\end{align}
The corresponding projection operators are
\begin{multline}
  \left(P_\alpha^{\,c\,}\right)=
  \left[\,(1-n_{i\uparrow})(1-n_{j\uparrow}),\ (1-n_{i\uparrow})n_{j\uparrow},\ \right. \\
  \left. n_{i\uparrow}(1-n_{j\uparrow}),\ n_{i\uparrow}n_{j\uparrow}\,\right]^T,
\end{multline}
and
\begin{multline}
  \left(P_\alpha^{\,s\,}\right)=
  \left[\,
    n_{i\uparrow}(1-n_{i\downarrow})n_{j\uparrow}(1-n_{j\downarrow}),\ 
  \right. \\
    n_{i\uparrow}(1-n_{i\downarrow})(1-n_{j\uparrow})n_{j\downarrow},\ \\
    (1-n_{i\uparrow})n_{i\downarrow}n_{j\uparrow}(1-n_{j\downarrow}),\ \\
  \left. 
    (1-n_{i\uparrow})n_{i\downarrow}(1-n_{j\uparrow})n_{j\downarrow}\,
  \right]^T,
\end{multline}
with the upper indices ($c,s$) referring to charge and spin qubits (respectively). Subsequently, the transfer operators are given by 
\begin{equation}
  \left(\,T_{\alpha\alpha'}^{\,c}\right)=
  \left(
    \begin{array}{cccc}
      1 & c_{j\uparrow} & c_{i\uparrow} & c_{j\uparrow}c_{i\uparrow} \\
      c_{j\uparrow}^\dagger & 1 & c_{j\uparrow}^{\dagger}c_{i\uparrow} & c_{i\uparrow} \\
      c_{i\uparrow}^\dagger & c_{i\uparrow}^{\dagger}c_{j\uparrow} & 1 & -c_{j\uparrow} \\
      c_{i\uparrow}^{\dagger}c_{j\uparrow}^{\dagger} & c_{i\uparrow}^{\dagger} & 
      -c_{j\uparrow}^{\dagger} & 1 \\
    \end{array}
  \right), 
\end{equation}
and
\begin{equation}
  \left(\,T_{\alpha\alpha'}^{\,s}\right)=
  \left(
    \begin{array}{cccc}
      1 & S_j^+ & S_i^+ & S_j^+{}S_i^+ \\
      S_j^- & 1 & S_j^-{}S_i^+ & S_i^+ \\
      S_i^- & S_i^-{}S_j^+ & 1 & S_j^+ \\
      S_i^-{}S_j^- & S_i^- & S_j^- & 1  \\
    \end{array}
  \right), 
\end{equation}
with the spin operators $S_i^+=c_{i\uparrow}^\dagger{}c_{i\downarrow}$,  $S_i^-=c_{i\downarrow}^\dagger{}c_{i\uparrow}$. 
Substituting the above expressions into Eq.\ (\ref{rhostar}) we get
\begin{equation}
  \label{rhoalpx}
  (\,\rho_{\alpha\alpha'}^X)=
  \left(\begin{array}{cccc}
    u_+^X & 0       & 0     & 0 \\
    0     & w_1^X   & z^X   & 0 \\
    0     & (z^X)^\star & w_2^X & 0 \\
    0     & 0       & 0     & u_-^X \\
  \end{array}\right)
  \ \ \ \ \text{for}\ X=c,s,
\end{equation}
where the total spin and charge conservation is imposed. The nonzero matrix elements in Eq.\ (\ref{rhoalpx}) are given by 
\begin{gather}
  u_+^c = \langle{(1-n_{i\uparrow})(1-n_{j\uparrow})}\rangle, \ \ \ \ 
  w_1^c = \langle{n_{i\uparrow}(1-n_{j\uparrow})}\rangle,
  \nonumber \\
  z^c = \langle{c^{\dagger}_{j\uparrow}c_{i\uparrow}}\rangle, \ \ \ \ 
  w_2^c = \langle{(1-n_{i\uparrow})n_{j\uparrow}}\rangle, 
  \nonumber \\
  u_-^c = \langle{n_{i\uparrow}n_{j\uparrow}}\rangle,  \label{uwzchar}
\end{gather}
for charge qubits, or by 
\begin{align}
  u_+^s &= \langle{
    n_{i\uparrow}(1-n_{i\downarrow})n_{j\uparrow}(1-n_{j\downarrow})
  }\rangle, \nonumber \\
  w_1^s &= \langle{
    n_{i\uparrow}(1-n_{i\downarrow})(1-n_{j\uparrow})n_{j\downarrow}
  }\rangle, \nonumber \\ 
  z^s &= \langle{S_j^+S_i^-}\rangle = 
  \langle{
    c^{\dagger}_{j\uparrow}c_{j\downarrow}c^{\dagger}_{j\downarrow}c_{j\uparrow}
  }\rangle, \nonumber \\
  w_2^s &= \langle{
    (1-n_{i\uparrow})n_{i\downarrow}n_{j\uparrow}(1-n_{j\downarrow})
  }\rangle,
  \nonumber \\
  u_-^s &= \langle{
    (1-n_{i\uparrow})n_{i\downarrow}(1-n_{j\uparrow})n_{j\downarrow}
  }\rangle,  \label{uwzspin}
\end{align}
for spin qubits. 

We use now the {\em concurrence} ${\cal C}$, as a~quantitative measure of quantum entanglement in the two-qubit subsystem $A$. The closed-form expression was derived by Wootters \cite{Woo98} and reads
\begin{equation}
\label{conc:def}
  \mathcal{C} = \max\left\{0,\sqrt{\lambda_1}-\sqrt{\lambda_2}-
  \sqrt{\lambda_3}-\sqrt{\lambda_4}\right\},
\end{equation}
where $\lambda_1\geqslant\lambda_2\geqslant\lambda_3\geqslant\lambda_4$ are eigenvalues  of the matrix product 
\begin{equation}
  \label{matprod}
  \rho\cdot (\sigma_i^y\otimes\sigma_j^y)\rho^* (\sigma_i^y\otimes\sigma_j^y)  
\end{equation}
with $\rho=(\,\rho_{\alpha\alpha'}^X)$ given by Eqs.\ (\ref{rhoalpx}), (\ref{uwzchar}), and (\ref{uwzspin}), and $\sigma_i^y$ ($\sigma_j^y$) being the second Pauli matrix acting on the qubit associated with $i$-th ($j$-th) lattice site. It was also shown in Ref.\ \cite{Woo98} that the entanglement of formation for a~pair of qubits was uniquely determined by ${\cal C}$, namely 
\begin{equation}
  E_f=-\xi\log_2\xi-(1-\xi)\log_2(1-\xi),
\end{equation}
where $\xi=(1+\sqrt{1-\mathcal{C}^2})/2$. In fact, ${\cal C}$ can be interpreted as a~distance between a~given quantum state and the nearest separable state \cite{Woo01}. For these reasons, ${\cal C}$ can be used to quantify the entanglement of two qubits instead of $E_f$, and is hereinafter called {\em pairwise entanglement}. 

The eigenvalues of the matrix defined by Eq.\ (\ref{matprod}) can be written as
\begin{align}
  \tilde{\lambda}_1 &= \tilde{\lambda}_4 = u_+^X u_-^X, \nonumber \\ 
  \tilde{\lambda}_{2,3} &= \left(\sqrt{w_1^Xw_2^X}\pm|{z^X}|\right)^2,
\end{align}
where $\tilde{\lambda}_1,\dots,\tilde{\lambda}_4$ are yet unsorted. After some straightforward steps, Eq.\ (\ref{conc:def}) leads to
\begin{equation}
  \label{conc:valx}
  {\cal C}=2\max\left\{0,|z^X|-\sqrt{u_+^X u_-^X}\right\},
\end{equation}
where the relevant correlation functions are given by Eq.\ (\ref{uwzchar}) for $X=c$, or by Eq.\ (\ref{uwzspin}) for $X=s$. The correlation functions $w_1^X$ and $w_2^X$ are absent in Eq.\ (\ref{conc:valx}) as they contribute to $\sqrt{\lambda_1}-\sqrt{\lambda_2}-\sqrt{\lambda_3}-\sqrt{\lambda_4}$ if and only if  $\sqrt{\lambda_1}-\sqrt{\lambda_2}-\sqrt{\lambda_3}-\sqrt{\lambda_4}<0$.  

Generalizations of ${\cal C}$ for multifermionic states, appearing when larger subsystem $A$ is under consideration, are also possible \cite{Eck02}. Such generalizations are, however, beyond the scope of this paper. We focus here on the pairwise entanglement, mainly because the form of Eq.\ (\ref{conc:valx}) makes it possible to be determined with no significant computational costs once standard (spin or charge) correlation functions are calculated.

\begin{figure}
%\centerline{\includegraphics[width=0.9\linewidth]{conc-half}}
\centerline{\includegraphics[width=0.9\linewidth]{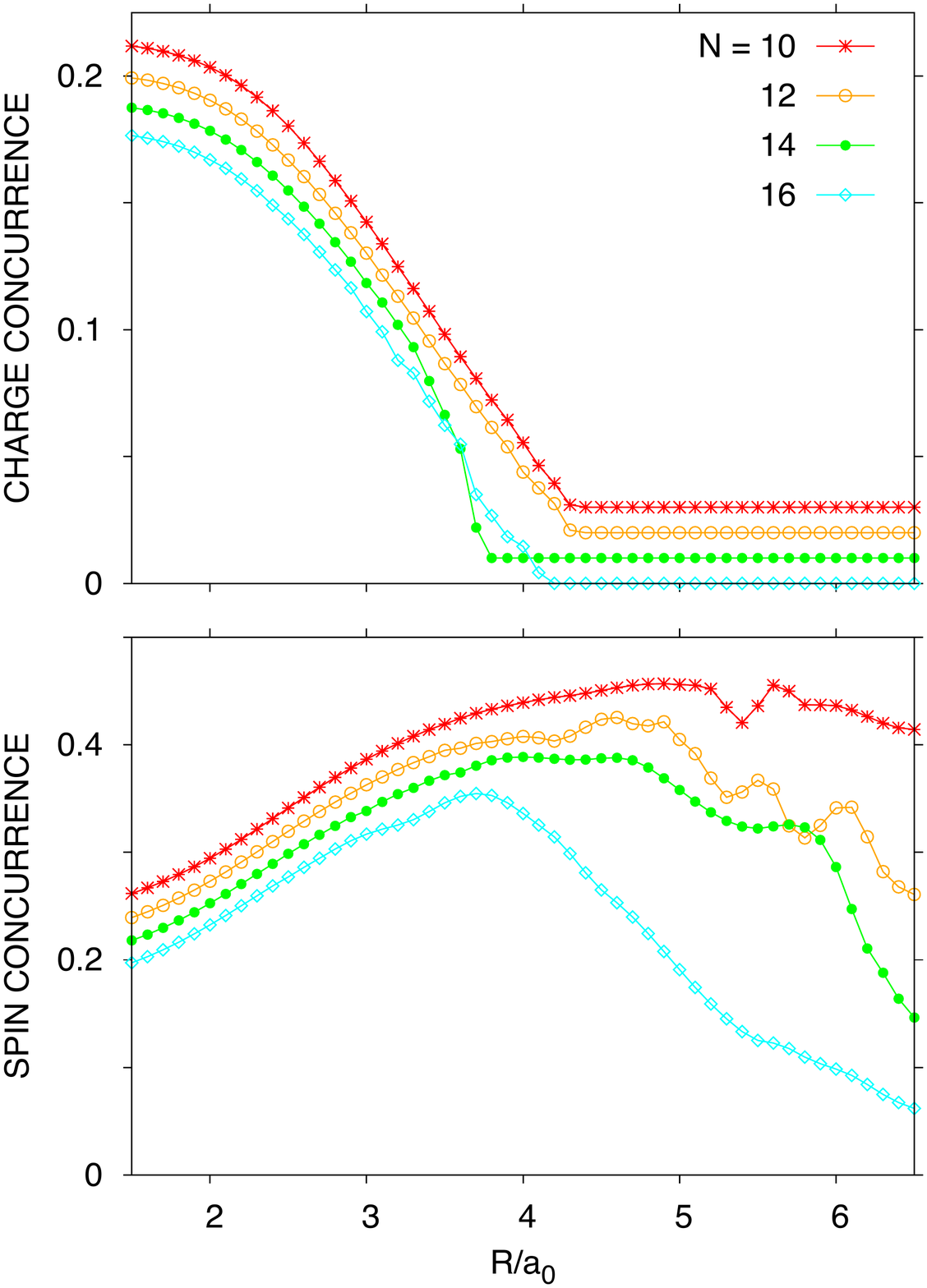}}
\caption{ \label{fig:conc-half}
  Nearest-neighbor 
  pairwise entanglement calculated from Eq.\ (\ref{conc:valx}) for charge 
  (top panel) and spin (bottom panel) degrees of freedom; 
  $N=N_{\rm el}=10-16$. 
  The vertical shifts, applied to the datasets in both panels, are: 
  $0$ ($N=16$), $0.01$ ($N=14$), $0.02$ ($N=12$), and  $0.03$ ($N=10$). 
  Lines are guides for the eye only.
%  {\color{red}\sf\dots 
%    Zmieniony dolny panel - spr. odwolania w tekscie! \dots
%  }
}
\end{figure}

\begin{figure}
%\centerline{\includegraphics[width=0.9\linewidth]{conc-quart}}
\centerline{\includegraphics[width=0.9\linewidth]{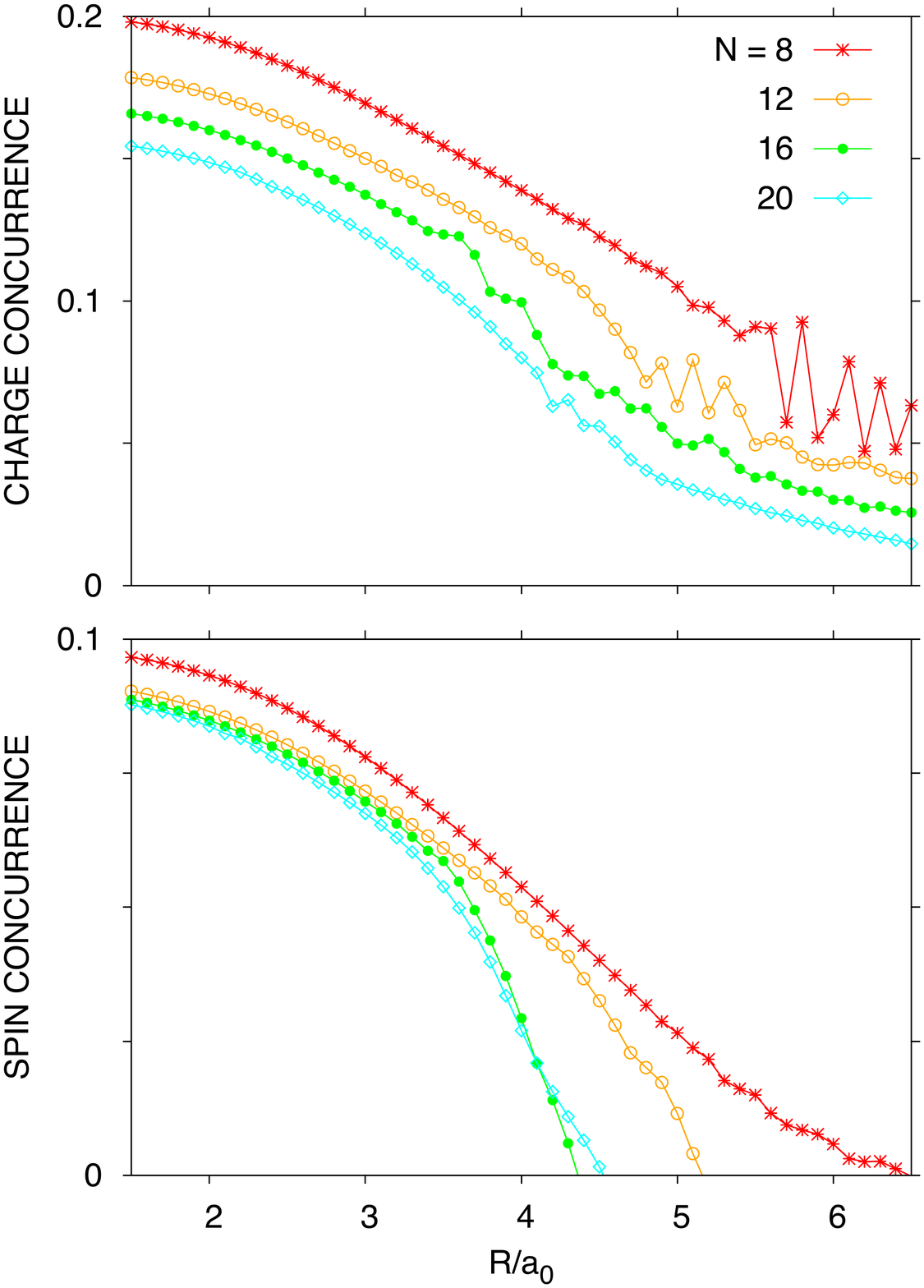}}
\caption{ \label{fig:conc-quart}
  Same as Fig.\ \ref{fig:conc-half}, but for $N=8-20$ and $N_{\rm el}=N/2$. 
  The vertical shifts, applied to the datasets in the top panel only, 
  are: $0$ ($N=20$), $0.01$ ($N=16$), $0.02$ ($N=12$), and  $0.03$ ($N=8$). 
}
\end{figure}

Our numerical results for the pairwise entanglement are presented in Figs.\ \ref{fig:conc-half} and \ref{fig:conc-quart}, constituting the central findings of this paper. The presentation is limited to the cases when sites $i$ and $j$ are the nearest neighbors in a~linear chain (i.e., $j=i\pm{}1\,$mod$\,N$), as we have found that ${\cal C}=0$ for more distant neighbors \cite{cdistfoo}. Also, correlation functions in Eq.\ (\ref{conc:valx}) are averaged over the reference site number $i$, to reduce the finite-precision effects. 

For $N_{\rm el}=N$, we have ${\cal C}=0$ for the charge degrees of freedom (so-called {\em charge concurrence}) provided that $R/a_0\gtrsim{}4$ (see top panel in Fig.\ \ref{fig:conc-half}). 
In such a~range, the separability of a~quantum state in the position representation, appearing if a~bipartite subsystem ($A$) is selected in a~fixed-spin sector, justifies the above-used notion of a~fully-reconstructed Mott insulator for a~finite $N$ (see  Sec.\ III and Refs.\ \cite{Ryc04,Ste11}). 
For smaller $R$, stronger charge fluctuations manifest itself via nonzero charge concurrence, providing an insight into the nature of a~partly-localized quantum liquid. Remarkably, spin concurrence (see bottom panel in Fig.\ \ref{fig:conc-half}) is positive for any considered $N$ and $R/a_0$, reaching the maximum in the crossover range of $4\lesssim{}R/a_0\lesssim{}5$. For this reason, spin concurrence still can be considered as a~prospective signature of the Mott transition. 

For $N_{\rm el}=N/2$ the evolution of the pairwise entanglement with $R/a_0$ (see Fig.\ \ref{fig:conc-quart}) is significantly different than for the $N_{\rm el}=N$ case. For each $N$, we have a~nonzero charge concurrence for any considered $R/a_0$ (top panel in  Fig.\ \ref{fig:conc-quart}), whereas the spin concurrence (bottom panel) indicates a~finite-system version of the Mott transition \cite{lanczfoo}. In brief, for spin degrees of freedom ${\cal C}=0$ for $R>R_\star(N)$, where the nodal value $R_\star(N)$ corresponds to $|z^s|-\sqrt{u_+^s{}u_-^s}=0$,  and is determined numerically via least-squares fitting of a~line to the actual datapoints for a~given $N$ (see the last column in Table \ref{tab:rstara}). For larger $N$, the values of $R_\star(N)$ are close to $R_c\approx{}4.2\,a_0$, determined in Sec.\ III via the finite-size scaling for $\Delta{}E_C$.  

Additionally, we observe that the nearest-neighbor pairwise entanglement may indicate the decoupling of charge and spin degrees of freedom in correlated systems. This happens for $R/a_0\gtrsim{}4$ in the $N_{\rm el}=N$ case (as well as for $R>R_\star(N)$ in the $N_{\rm el}=N/2$ case), where  nonzero spin (charge) concurrence is accompanied by vanishing charge (spin) concurrence. In the above-mentioned ranges, quantum states of a~bipartite subsystem ($A$) are entangled when spin, but separable when charge degrees of freedom are under consideration ($N_{\rm el}=N$), or vice versa ($N_{\rm el}=N/2$). Such a~decoupling should be distinguished from charge-spin separation predicted to appear in 1D systems showing the Luttinger liquid phase \cite{Sac13,Lie68,Voi93,Ber00}. Our numerical examples also show there is  no one-to-one relation between the decoupling and the Mott transition, as the format can be identified either for the system showing the crossover behavior only or undergoing the Mott transition (in the large-$N$ limit). 
Finally, the pairwise-entanglement analysis presented above allows one to identify noticeably different behavior of charge and spin correlation functions without referring to their asymptotic behavior, as previously proposed in the literature (see the first paper in Ref.\ \cite{Ste11} and Refs.\ \cite{Sac13,Gal16}).

%%%%%%%%%%%%%%%%%%%%%%%%%%%%%%%%%%%%%%%%%%%%%%%%%%%%%%%%%%%%%%%%%%%%%%%%

\section{Conclusions}
In this study we have demonstrated that pairwise entanglement, quantified by the fermionic concurrence determined separately for charge and spin degrees of freedom, can serve as a~convenient indicator for a~finite-system version of the Mott transition. In particular, standard finite-size scaling estimates of the Mott transition for linear chains of hydrogenic-like atoms are revisited utilizing the exact diagonalization -- ab initio metod (EDABI) at the half and the quarter electronic filling. In the latter case, we find that not merely the charge gap indicates the transition for $1/N\rightarrow{}0$ (with $N$ being the number of atoms) at the interatomic distance $R_c\approx{}4.2\,a_0$ (where $a_0$ denotes the Bohr radius), but also the spin concurrence vanishes for a~finite $N$ at $R_\star(N)$ taking the values relatively close to $R_c$. Charge concurrence remains nonzero in both the metallic and the insulating phases, signaling a~decoupling between charge and spin degrees of freedom. 

At the half filling the Mott transition is not observed. Instead, the crossover from a~partly localized quantum liquid to a~fully-reconstructed Mott insulator occurs. The charge concurrence vanishes at the crossover point, where the spin concurrence shows a~broad peak; the latter remains nonzero in the entire parameter range.

It is worth to stress here that calculations of the fermionic concurrence, employed in this paper, generate essentially no extra computational costs, as the concurrence is determined solely via pairwise ground-state correlation functions for charge or spin degrees of freedom, without referring to the dynamical properties such as the optical or {\em dc} conductivity. 
The analytic relations between entanglement and the correlation functions [see Eq.\ (\ref{conc:valx})] are also derived in this paper. 

A~striking feature is that the analysis holds true regardless how the exact (or approximated) ground state is obtained, making it possible to employ the entanglement-based signatures of the Mott transition when discussing a~generic correlated-electron system. 
What is more, Eq.\ (\ref{conc:valx}) also applies when correlation functions on its right-hand side are determined using more powerful numerical techniques, such as DMRG or QMC. 
For these reasons, 
we believe the approach we propose may shed new light on various open problems in the field, such as the long-standing metallization of solid hydrogen \cite{Kad14}, or the recently-raised phase diagram of the repulsive Hubbard model on a~honeycomb lattice \cite{spnlqd}. 

%$\Rightarrow$
%{\sf JESZCZE: problem ze zbieznoscia przerwy energet/spinowej generuje
%potrzebe znalezienia bardziej efektywnych wyznacznikow MIT, spin-conc. moze
%byc rozwiazaniem!}

\section*{Acknowledgements}
I thank Andrzej Biborski, Andrzej K\k{a}dzielawa, and Prof.\ J\'ozef Spa{\l}ek for many discussions,  and Yichen Huang for pointing out the role of asymptotic properties of the reduced density matrix. 
The work was supported by the National Science Centre of Poland (NCN) 
via Programme SONATA BIS, Grant No.\ 2014/14/E/ST3/00256. 
Computations were performed using the PL-Grid infrastructure.

%%%%%%%%%%%%%%%%%%%%%%%%%%%%%%%%%%%%%%%%%%%%%%%%%%%%%%%%%%%%%%%%%%%%%%%%
%%%%%%%%%%%%%%%%%%%%%%%%%%%%%%%%%%%%%%%%%%%%%%%%%%%%%%%%%%%%%%%%%%%%%%%%

%%%%%%%%%%%%%%%%%%%%%%%%%%%%%%%%%%%%%%%%%%%%%%%%%%%%%%%%%%%%%%%%%%%%%%%%
%%%%%%%%%%%%%%%%%%%%%%%%%%%%%%%%%%%%%%%%%%%%%%%%%%%%%%%%%%%%%%%%%%%%%%%%


\begin{thebibliography}{}
\bibitem{spingapfoo}
Strictly speaking, finding out whether the insulating phase is of the Mott type also requires calculation of the so-called spin gap, $\Delta{}E_S=E_G^{S_z=1}-E_G^{S_z=0}$ (where $E_G^{S_z}$ denotes the lowest eigenenergy in the subspace with a~total $z$-th component of spin $S_z$);  $\Delta{}E_C>0$ accompanied by $\Delta{}E_S\rightarrow{}0$ for $1/N\rightarrow{}0$ indicates the Mott phase. Although the answer may not be obvious in a~more general situation (see Ref.\ \cite{spnlqd}), a~lack of long-range spin order in 1D systems manifests itself by a~fast decay of $\Delta{}E_S$ with $N$ for atomic chains described by the Hamiltonian  $\hat{\cal H}(\alpha,R)$ (\ref{hamchain}) for all values of $R$. The numerical demonstrations are provided in Refs.\ \cite{Ryc04,Rycphd}.
\bibitem{cdistfoo}
This can be attributed to generic asymptotic behavior of the reduced density matrix [see Eq.\ (\ref{rhoalpx})], for which $\rho_{\alpha\alpha'}^X\approx{}(\frac{1}{2}{n})^2\delta_{\alpha\alpha'}$ for $X=c$, or $\rho_{\alpha\alpha'}^X\approx{}(\frac{1}{2}{n}-d)^2\delta_{\alpha\alpha'}$ for $X=s$, since the relevant operator-fluctuations in Eqs.\ (\ref{uwzchar}) and (\ref{uwzspin}) become uncorrelated with growing the distance between sites $i$ and $j$. Analogous features for spin models were discussed in Ref.\ \cite{Hua14}.
\bibitem{lanczfoo}
We have intentionally limited the number of Lanczos iterations, starting (in any case) from a~randomly-chosen vector in the Fock space, to relatively small value of $m=30$. This allows us to visualize how GS correlation functions converges with the system size: Notice that the oscillations of ${\cal C}$ as a~function of $R$, visible for $N=8$ and $12$ in Fig.\ \ref{fig:conc-quart}, are strongly suppressed for larger $N$.
\end{thebibliography}
\end{document}